\documentclass{article}
\usepackage{arxiv}
\usepackage{graphicx}
\usepackage{commath}
\usepackage{bm}
\usepackage[T1]{fontenc}
\usepackage{hyperref} 
\usepackage{url}      
\usepackage{booktabs}   
\usepackage{amsfonts} 
\usepackage{nicefrac}    
\usepackage{cite}
\usepackage{color}
\usepackage{array}
\usepackage{mdwmath}
\usepackage{mdwtab}
\usepackage{amssymb,amsmath,amsthm,mathtools}
\usepackage{enumerate}
\usepackage{setspace}
\usepackage{cleveref}
\usepackage{float}
\usepackage[ruled]{algorithm2e}

\title{On the Use of Reinforcement Learning for Attacking and Defending Load Frequency Control}
\date{} 					

\author{{Amr S.~Mohamed}\\
	Department of Electrical Engineering\\
	University of Toronto\\
	Toronto, ON M5S 3G4, Canada \\
	\texttt{amr.mohamed@mail.utoronto.ca} \\
    \And
    {Deepa~Kundur}\\
	Department of Electrical Engineering\\
	University of Toronto\\
	Toronto, ON M5S 3G4, Canada \\
    \texttt{dkundur@ece.utoronto.ca}}

\begin{document}
\maketitle

\begin{abstract}
The electric grid is an attractive target for cyberattackers given its critical nature in society.
With the increasing sophistication of cyberattacks, effective grid defense will benefit from proactively identifying vulnerabilities and attack strategies. 
We develop a deep reinforcement learning-based method that recognizes vulnerabilities in load frequency control, an essential process that maintains grid security and reliability.
We demonstrate how our method can synthesize a variety of attacks involving false data injection and load switching, while specifying the attack and threat models -- providing insight into potential attack strategies and impact.
We discuss how our approach can be employed for testing electric grid {vulnerabilities}. Moreover our method can be employed to generate data to inform the design of defense strategies and develop attack detection methods. 
For this, we design and compare a (deep learning-based) supervised attack detector with an unsupervised anomaly detector to highlight the benefits of developing defense strategies based on identified attack strategies.
\end{abstract}


\section{Introduction} \label{sec:intro}

The electrical power grid is evolving to provide enhanced availability, efficiency, and reliability of electricity through an increased reliance on information and communication technologies~\cite{national2007netl}.
This modernization has and will continue to introduce new and critical cybersecurity vulnerabilities~\cite{lee2010guidelines}. 
If exploited by cyberattackers, the resulting damage can have devastating consequences to the welfare of society, including economic loss, injury, and loss of life~\cite{baumeister2010literature}. 

Recent cyberattacks, such as the 2015 Ukrainian grid attack, left 225,000 people without electricity for hours by sabotaging operator workstations, wiping system files, flooding phone lines, and disabling backup power supplies to bring down the grid and impede its restoration~\cite{sullivan2017cyber}. Such attacks have shown to exhibit prior system knowledge on the part of the attacker, stealth and a high degree of sophistication. We assert that without a reasonable understanding of attacker resources and strategies, electric grid defense is limited to taking a reactive stance leaving the defender at a fundamental disadvantage and to be more easily bypassed~\cite{yu2015blind, zhang2020detecting, lakshminarayana2020data, zhang2020market, mukherjee2022data}. Hence, in this paper we consider a more proactive perspective of defense.

{
To detect common forms of data corruption attacks, power systems have traditionally relied on bad data detection (BDD) methods, which were originally developed to detect highly corrupt measurements (often stemming from telemetry error). BDD methods use historical data sets, statistical approaches, and approximate system models to flag abnormal measurements and are thus limited to detecting more naively constructed cyberattacks. Specifically, these methods, used as a reactive form of defense, fail to detect attacks that either exploit model inaccuracy or  are intentionally crafted such that their distribution is similar to that of the historical system data~\cite{sayghe2020survey}. To address these limitations, recent research on attack detection has leveraged data-driven methods such as machine learning  \cite{ozay2015machine, he2017real, james2018online} and more recently deep learning \cite{chen2017novel, abbaspour2019resilient}. Studies have shown how these methods are more effective than traditional BDD especially in detecting false data injection (FDI) in the context of state estimation or load frequency control in power systems.
} 


{
Nevertheless, these data-driven methods are typically evaluated using attacks that are randomly generated or crafted using a simple library of templates, which we argue fail to assess their performance against more realistic attacks that are complex in nature and targeted based on knowledge of system vulnerabilities and dynamics. Thus, we assert that the effectiveness of learning-based methods against such attacks remains largely untested. Further, we believe that a proactive stance is needed to identify and address vulnerabilities. 
}


One strategy, which is the focus of this paper, is to synthesize novel attacks (via intelligent attacker modelling) to inform defense development pre-emptively. {The goal of attack synthesis, here, would be to} provide insight into attacker strategies to forecast security requirements, appropriately reinforce grid defenses, and improve situational awareness.
{Computationally-based attack synthesis approaches include generative adversarial networks (GAN) (e.g., \cite{liu2022gan}), optimization (including reachability frameworks, e.g., \cite{esfahani2010cyber}), and reinforcement learning (RL). 
GANs learn known attack patterns to synthesize additional attack realizations; however, prior knowledge of attacks is required, which limits their ability to synthesize new unknown attacks. 
Optimization-based methods are model-based, requiring accurate system models to synthesize effective attacks, and strong assumptions on the system and/or attack models. 
{In contrast,} RL agents can learn new attacks with zero to little prior knowledge of the system and attacks. Further, RL is data-driven, relying on partial system observations, which can involve complex dynamics and inter-dependencies.
}

{
Given these advantages, RL is increasingly being applied to electric grid attack and defense.
Various studies have constructed (Q-learning) RL agents to synthesize~\cite{yan2016q, ni2017reinforcement, wang2020coordinated, paul2018study, ni2019multistage} and develop defense strategies~\cite{paul2018study, ni2019multistage} against line-switching attacks that exploit how sudden changes in grid topology can lead to cascading failures and blackout. 
Additionally, RL has been applied to the synthesis of FDI attacks in power systems~\cite{chen2018evaluation}, where the RL agent mimics a virus in a compromised power substation attempting to induce voltage sags in the system. 
A multi-agent RL approach has also been proposed in~\cite{abianeh2021vulnerability} to synthesize FDI attacks that can bypass attack detection methods in DC microgrids.
}

In this work, we extend RL research for attack synthesis against the electric grid to load frequency control (LFC). Frequency deviation negatively impacts grid operation, security, and reliability~\cite{mohan2020comprehensive}, and can potentially result in equipment damage, load performance degradation, transmission line overload, generation loss, and grid instability~\cite{wu2017resonance}. 
{
Due to its critical role in maintaining nominal grid frequency, LFC is a valuable target of cyberattacks.}
As such, the security of LFC has been the subject of a growing body of research papers~\cite{mohan2020comprehensive}.


{
In this research, we build on this literature to demonstrate the utility of RL in replicating known attacks and synthesizing new realizations against LFC. We show how our RL paradigm holistically explores the attack space to expose possible attacker strategies to help specify attack requirements and verify attack/threat model assumptions to improve electric grid defense. Hence, in this paper,
\begin{enumerate}
    \item we employ RL, for the first time, in the synthesis of attacks against LFC, training RL agents to execute FDI and load switching attack strategies. 
    To the best of our knowledge, our study is the first to apply RL to dynamic power system cyber-physical security. Unlike previous studies that focused on the RL agent's effect on power flow and state estimation computations, we investigate its influence on the power system's dynamics and validate results empirically. Additionally, we provide novel RL reward function formulations that serve as templates for facilitating the training of RL agents against LFC. 
    {The reward functions can guide future research by serving as a blueprint for rewarding and training RL agents to relieve or induce stress on the power system by deviating from its nominal states.}
    \item we present and discuss various novel benefits to utilizing RL for cyber-physical security, including replicating known attacks, exploring the attack space, revealing potential attack strategies, specifying attack/threat model assumptions, and developing proactive defense strategies.
    \item we harness the RL generated data to train a supervised learning-based attack detector with a long short-term memory (LSTM) neural network, and compare it with the state-of-the-art unsupervised anomaly detection based on autoencoders to demonstrate the benefits of RL-based attack synthesis for defense. In this way, we demonstrate the utility of RL attack synthesis in improving detection-based mitigation.
\end{enumerate}
}

This paper is structured as follows. In Section~\ref{sec:background} we review LFC and {survey cyberattacks targeting LFC systems}. {We provide relevant background on RL.} Section~\ref{sec:threatModel} discusses the threat model that we consider. RL agent and attack detection methodologies are developed in Section~\ref{sec:method} followed by presentation of RL training and attack detector performance results with discuss on benefits and challenges in Section~\ref{sec:results}. Section~\ref{sec:conclusion} provides concluding remarks.

\section{Background} \label{sec:background}

Our research focuses on the use of RL to generate attacks against LFC and subsequently leveraging the RL-generated attacks to devise defense strategies. 
{Hence, in this section, we introduce LFC and its vulnerabilities to cyberattacks and summarize RL.}

\subsection{Frequency Control and Protection} \label{subsec:LFC}

{
LFC maintains power balance and grid frequency consisting of primary, secondary, and tertiary control levels \cite{kundur2007power}. The primary level employs droop-governor control to regulate frequency while the secondary level uses automatic generation control (AGC) to regulate the net interchange of power. Tertiary control provides additional frequency support mechanisms by restoring power reserve. Failure to regulate frequency can cause frequency protection devices (including ANSI 81U/O/R) to isolate power system equipment to protect them from damage sustained due to operation at abnormal frequency resulting in unwanted system reduction.

The implementation of LFC over a wide-area network with open communication protocols and minimal human supervision increases its cyberattack surface.  
Interference of LFC operation is possible by exploiting a variety of vulnerabilities in insecure legacy electric grid networks, open communication protocols and operating systems, as well as introducing malware through infected emails or USBs, performing supply-chain attacks, or capitalizing on disgruntled insiders~\cite{flick2010securing, glenn2016cyber, veitch2013microgrid}. 
These cyberattacks often aim to compromise critical measurement signals to ultimately destabilize the grid.
}

This work focuses on cyberattacks aimed at the unwanted triggering of frequency protection relays in power grids, which can initiate sudden power imbalance leading to grid instability, cascading failure, and blackout.
{
Previous studies highlight the significance of investigating these attacks, showing that attackers can modify loads~\cite{wu2017resonance, hammad2015tuning, hammad2017class, kabir2021two} or tie-line power~\cite{wu2017resonance}, inject disturbances to automatic generation control~\cite{esfahani2010cyber}, or corrupt microgrid synchronization~\cite{mohamed2021false} to trigger protection devices and/or destabilize the grid.
}
{
While these authors applied strong assumptions and/or knowledge of the system dynamics in the development of cyberattacks, in this paper we demonstrate that RL can synthesize attacks with zero to little prior knowledge. 
}

\subsection{Reinforcement Learning} \label{subsec:RL}

In RL, an agent is trained through a process of trial-and-error to achieve optimal decisions or strategies in an environment of which the agent has zero to little prior knowledge~\cite{sutton2018reinforcement}. 
Training an RL agent to attack the grid can yield novel unforeseen insight into system vulnerabilities and attack strategies. Establishing an RL problem within the context of synthesising attacks requires defining the \textit{environment} (representing the cyber-physical system), and specifying what \textit{actions} (representing attacks) the agent can execute in the environment and what environmental states it can \textit{observe} and make decisions based on. 
A \textit{reward function} is formulated to steer the agent into taking actions that achieve the attack goals. 

The agent contains two components: a \textit{policy} and a \textit{learning algorithm}.
The goal of the policy is to map environment observations to actions that maximize rewards. 
The policy can involve an actor, critic, or actor-critic function approximators. 
An actor $\pi : S \rightarrow A$ maps environment observations $S$ to actions $A$. A critic $Q : (S,A) \rightarrow R$ maps action-observation pairs to (predicted) discounted cumulative long-term rewards $R$.
The learning algorithm continuously updates the policy to find the optimal policy.
Learning happens in \textit{episodes}: simulations that expire after the RL agent achieves a certain goal or a maximum simulation length.


{
In this work, we employ deep deterministic policy gradient (DDPG) RL as it is the \textit{simplest} learning algorithm compatible with continuous actions and observations -- to highlight the feasibility of implementing our work offensively as a cyberattacker or defensively for electric grid cyber-physical security.
}

\section{Threat Model} \label{sec:threatModel}

We train RL agents to execute attacks against LFC that directly lead to protection tripping and loss of generation. 
Offensively, the agent can be deployed by a cyberattacker during a cyber breach in a man-in-the-middle or FDI attack, or be packaged within (malicious) software that is uploaded to a {control device}. 
We assume the following salient components of the threat model:
\begin{itemize}
    \item The electric grid exhibits oscillatory eigenmodes that can be leveraged. 
    Generators' mechanical construction gives rise to their own eigenmodes. The existence of inter-area oscillatory eigenmodes is evident in most multi-machine power systems~\cite{hammad2017class}.

    \item The attacker can perform one of the following actions:
    \begin{enumerate}
    \item corrupt frequency (sensor) measurements to the control center~\cite{khalaf2018joint},
    \item corrupt generation control signals~\cite{mohamed2021false, esfahani2010cyber},
    \item corrupt tie-line power measurements~\cite{wu2017resonance, khalaf2018joint}, or
    \item compromise loads~\cite{wu2017resonance, hammad2015tuning, hammad2017class, kabir2021two}. 
    \end{enumerate}
    {For detailed information on threat models and attack execution, please refer to the cited papers.}
    \item The attacker can observe the grid frequency.
    Since frequency is a global state of the electric grid, the attacker can easily measure the grid frequency~\cite{wu2017resonance}, and compute its derivative (rate of change) and/or time-integral.
    
    \item The attacker is not assumed to have any knowledge of the system; hence, we initialize the RL agent with zero knowledge of its environment.
\end{itemize}

\section{Method} \label{sec:method}

To the best of our knowledge, our work is the first to apply the methods presented in this section, including the dynamical power system model and machine learning algorithms, within an RL framework to attack and defend LFC. We first model LFC in the RL environment and develop the RL agents. Next, we construct the supervised attack detector and unsupervised anomaly detector. 
{Numerical details of the algorithms are addressed in Section \ref{sec:results}}.

\subsection{System Architecture} \label{subsec:architecture}

We apply the swing equation~\cite{kundur2007power} to model LFC.  
The following state-space system expresses the linear load-frequency dynamics: 
\begin{equation} \label{eq:stateSpace}
    \bm{\Dot{x}} = \bm{A} \bm{x} + \bm{B} \bm{u} + \bm{W} \bm{p}
\end{equation}
where the state is
\begin{equation}
    \bm{x} = \begin{bmatrix} \Delta e & \Delta P_g & \Delta P_m & \Delta \omega & \Delta \Hat{\omega} & \Hat{\Dot{\omega}} \end{bmatrix}^T,
\end{equation}
{$e$ is governor-droop control signal,
$P_{g}$, the governor output, $P_{m}$, the mechanical power, $\omega$,  the system frequency, $\hat{\omega}$, the frequency measurement, and $\hat{\dot{\omega}}$, the rate of change of frequency measurement.}

The input vectors $\bm{u}$ and $\bm{p}$ represent the inputs to the systems during normal operation and attacks, respectively. The input vector:
\begin{equation}
    \bm{u} = \begin{bmatrix} \Delta P_L & \Delta P_{tie} \end{bmatrix}^T
\end{equation}
{includes change in the demand, $P_L$, and tie-line power, $P_{tie}$, if any.} The attack vector: 
\begin{equation}
    \bm{p} = \begin{bmatrix} p_1 & p_2 & p_3 & p_4 \end{bmatrix}^T
\end{equation}
includes actions the attacker can execute as enumerated in our threat model, including 
corrupting frequency measurements to the control center ($p_1$), 
corrupting generation control signals ($p_2$),
corrupting tie-line power measurements ($p_3$),
and compromising load switching ($p_4$). 
The state matrices in (\ref{eq:stateSpace}) are elaborated in the Appendix.

Frequency relays protect generators from damage sustained during operation in unsafe frequency by ensuring the frequency and its rate of change are within a safe set 
\begin{equation} \label{eq:safeset}
    \mathcal{S} = \{ (\hat{\omega}, \hat{\dot{\omega}}) : \text{UF} \leq \hat{\omega} \leq \text{OF}, \abs{\hat{\dot{\omega}}} \leq \text{ROCOF} \}
\end{equation}
Frequency relay functions include under-(UF), over-(OF) and rate-of-change of-(ROCOF) frequency. 
Recommended settings per IEEE 1547~\cite{ieee1547} are listed in Table \ref{table:relaysettings}.

\begin{table}[hbt!]
\footnotesize
\centering
\caption{IEEE 1547 recommended relay settings for Category III}
\begin{tabular}[t]{lcc} 
\toprule
\textbf{Protection Function} & \textbf{Threshold} & \textbf{Clearing time}\\
\midrule
OF2 & $62.0$  Hz & $160$ ms\\
UF2 & $56.5$ Hz & $160$ ms\\
\midrule
ROCOF & 3 Hz/s & \\
\bottomrule
\label{table:relaysettings}
\end{tabular}
\end{table}

\subsection{RL for Attacking Load Frequency Control} \label{subsec:RLforLFC}

\begin{algorithm}
    \caption{DDPG algorithm for attacking LFC}\label{alg:ddpg}
    
    Initialize the mini-batch size $M$; actor and critic learning rates $\alpha_\theta, \alpha_\phi$ ; discount factor $\gamma$; target smooth factor $\tau$; episode length; training step length\;
    Define the action space $\mathcal{A}$, and noise distribution\;
    Initialize the critic $Q(S,A; \phi)$ and target critic $Q_t(S,A; \phi_t)$ neural networks with random parameters $\phi = \phi_t$\;
    Initialize the actor $\pi(S; \theta)$ and target critic $\pi_t(S; \theta_t)$ neural networks with random parameters $\theta = \theta_t$\;
    
    \For{each training episode}{        
        \For{each training step}{
            For the current observation $S = (\hat{\omega}, \hat{\dot{\omega}})$, select action $A = \pi(S; \theta) + N$ with noise $N$\;
            Execute action $A$ as an attack on the power system through one of the inputs in $\bm{p}$ (refer to (\ref{eq:stateSpace})). Observe the reward $R$ and the next observation $S'$\;
            Store the experience $(S,A,R,S')$ in the experience buffer\;
            Sample a random mini-batch of $M$ experiences $(S_i,A_i,R_i,S'_i)$ from the experience buffer\;
            \For{each sampled experience}{
                Calculate the value function target $y_i$\;
                \eIf{$S'_i$ is a terminal state}{
                    \begin{equation}
                        \footnotesize
                        y_i = R_i
                    \end{equation}
                }
                {
                    \begin{equation}
                        \footnotesize
                        y_i = R_i + \gamma Q_t(S'_i, \pi_t(S'_i; \theta_t); \phi_t)
                    \end{equation}
                } 
            }
            Compute the loss over mini-batch\;
            \begin{equation}
                \footnotesize
                L = \frac{1}{M} \sum_{i=1}^{M} (y_i - Q(S_i, A_i; \phi))^2
            \end{equation}
            Update critic parameters by minimizing over $L$\;
            \begin{equation}
                \footnotesize
                \phi \gets \phi - \alpha_\phi \frac{\partial L}{\partial \phi}
            \end{equation}
            Update actor parameters by descending policy gradient\;
            \begin{equation}
                \footnotesize
                \frac{\partial J}{\partial \theta} \gets  \frac{1}{M} \sum_{i=1}^{M} \frac{\partial}{\partial A} Q(S_i, A; \phi) \frac{\partial}{\partial \theta} \pi(S_i; \theta)
            \end{equation}
            \begin{equation}
                \footnotesize
                \theta \gets \theta - \alpha_\theta \frac{\partial J}{\partial \theta}
            \end{equation}
            End episode if $S \notin \mathcal{S}$, label $S$ as a terminal state. Store episode data\;

            Update the target actor and critic parameters periodically\;
            \begin{equation}
                \footnotesize
                \phi_t = \tau \phi + (1-\tau) \phi_t
            \end{equation}
            \begin{equation}
                \footnotesize
                \theta_t = \tau \theta + (1-\tau) \theta_t
            \end{equation}
    
        }
    }
\end{algorithm}

We assume that the attacker can observe the system frequency.
We model actions $\{p_1, p_2, p_3\}$, entailing corruption of communication, as continuous-value actions. 
In practice, the attacker's ability to inject an attack signal is limited due to physical constraints, constraints imposed by the communication protocol, or the need to avoid detection by bad data detectors.
Hence, we assume that the {attack vector $\bm{p}$ is bounded.}
For load switching, if the attacker compromises an aggregate load $P_{sw}$ and can switch on and off portions of the total load, then $p_4$ can also be modelled as continuous-value action, with $p_4 \in [0, P_{sw}]$.
If the attacker can only switch on and off the entire load, then $p_4$ is a discrete-value action $p_4 \in P_{sw} \times \{0,1\}$. 
{We discretize the DDPG RL actions in this case as follows:}
\begin{align}
    p_4 &= 0, &\text{\quad if  } p_4 < P_{sw}/2,\\
    &= P_{sw}, &\text{\quad if  } p_4 \geq P_{sw}/2 \nonumber
\end{align}

In Algorithm \ref{alg:ddpg}, we adapt DDPG to attack LFC.
Here, the RL agent observes the system state $S = (\hat{\omega}, \hat{\dot{\omega}})$ and influences the power system (\ref{eq:stateSpace}) by injecting an attack action $A$ through the input vector $\bm{p}$. 
The learning objective is for the RL agent to learn a policy $\pi : S \rightarrow A$ to force the states in $S$ outside the safe set $\mathcal{S}$ (\ref{eq:safeset}), effectively triggering a frequency protection device.
The attacker can then execute this policy to attack and destabilize a real system. 

{We collect all episodes in a dataset for training the supervised-learning attack detector. }
The DDPG neural network architecture is detailed in Table \ref{table:ddpgNNarchitectures} in the Appendix.
    
\subsection{Unsupervised Autoencoder-based Anomaly Detection} \label{subsec:autoencoder}

{Unsupervised machine learning methods detect potential cyberattacks by learning patterns and regularities in normal operation data and flagging anomalies. Due to the lack of labelled cyberattack datasets, unsupervised learning approaches, particularly autoencoder-based detectors, have seen increased application for attack detection~\cite{zhang2021deep}.

Autoencoders consist of a deep neural network -- partitioned into an encoder and decoder -- that is trained to reconstruct its input data. 
The encoder maps the input normal operation data to a compressed hidden representation} based on regularities in the data, and its decoder attempts to map this representation back to the input data.
When the trained autoencoder is applied to new data, a large variation between the data and the autoencoder's reconstruction indicates an anomaly in the data, which can then be classified as an attack.

\begin{figure}
    \centering
    \vspace{2mm}
    \includegraphics[width=0.6\columnwidth,trim={0 1cm 0 0},clip]{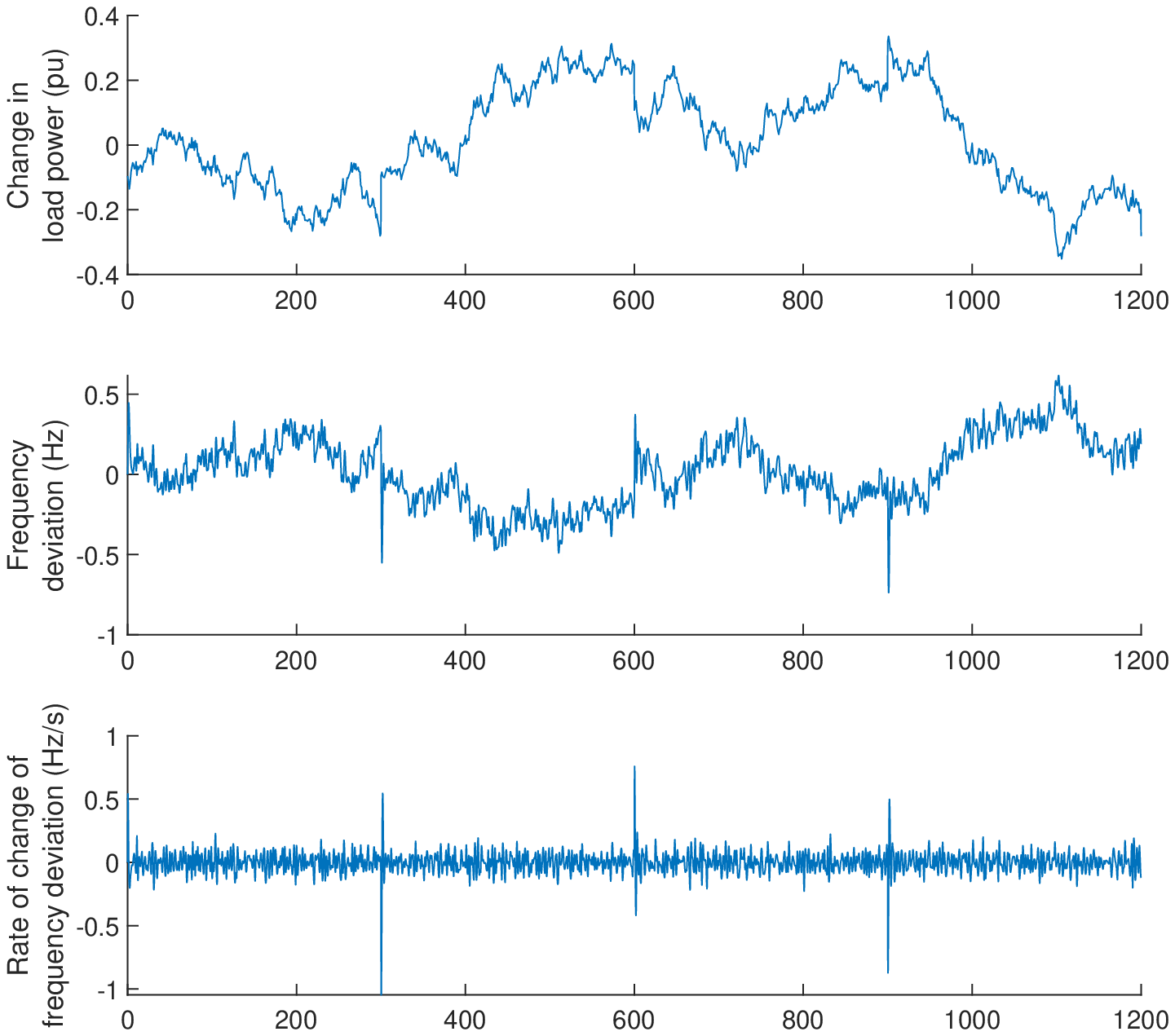}
    {\scriptsize Time (s)}
    \caption{Change in load power as a stochastic process with $\sigma_1 = (0.05/3)$ and $\sigma_2 = (0.2/3)$ to simulate normal grid operation data.}
    \label{fig:MGloadPower}
\end{figure}

To simulate normal operation data, we model the change in power demand as a random process following the work in~\cite{milano2013systematic, ferraro2016analysis}.
We model the change in demand as a 1 second-per-step random walk sampling from a Gaussian distribution $\mathcal{N}(0, \sigma_1^2)$ superimposed on a 5 minute-per-step random walk sampling from a Gaussian distribution $\mathcal{N}(0, \sigma_2^2)$. 
{Fig. \ref{fig:MGloadPower} illustrates the change in load power as simulated by this random process and the resulting frequency and rate of change of frequency deviation.}

We develop an autoencoder to detect anomalies in the time-series consisting of the governor-droop control signal $e$ and the frequency measurement $\hat{\omega}$. 
To prepare the training dataset, we run a long simulation of normal system operation and then we randomly crop $N$ portions.
The portions are sampled at 50 milliseconds per sample and have variable (time) length.
Each portion is a vector $X_i \in \mathbb{R}^{2 \times n_i}$ vector representing a time-series of ($e, \hat{\omega}$), where $n_i$ is the number of samples in the portion.

The overall autoencoder can be expressed as a mapping $f_{ae} : X_i \rightarrow \hat{X}_i$ between the input data $X_i$ and its reconstruction $\hat{X}_i$, where
$f_{ae}(X; \phi)$ is a neural network with parameters $\phi$.
We develop an LSTM neural network-based autoencoder, described in Table \ref{table:lstmNNarchitectures} of the Appendix, given the suitability of LSTM networks for time-series. 
Training the autoencoder seeks to minimize the mean square error between the data and their reconstructions:
\begin{equation}
    L_{MSE} = \frac{1}{N} \sum_{i=1}^{N} \frac{1}{2} (\hat{X}_i - X_i)^2
\end{equation}

Next, a threshold can be selected based on the maximum reconstruction error (seen in the validation set) to differentiate between normal and anomalous data. 
When the autoencoder is input new data, if the reconstruction error is below this threshold, the data is labelled normal; otherwise, anomalous. 

    
    
        
        

\subsection{Supervised attack detection using RL generated data} \label{subsec:supervisedDetector}

{Supervised methods are trained to distinguish between normal operation and attacks by training on labelled system data. 
To collect labelled data, we augment the normal operation dataset (used to train the autoencoder) with the RL-generated dataset} and label the data into 4 categories: (1) normal operation, and attacks that (2) do not trigger protection, (3) trigger under or over-frequency protection, and (4) trigger rate of change of frequency protection. 
Each record in the dataset includes a $X_i \in \mathbb{R}^{2 \times n_i}$ vector representing a time-series control signal and frequency measurement and a label $\ell_i \in \mathcal{C} = \{1,2,3,4\}$. 

We train the supervised attack detector to classify the data to their correct labels. 
The attack detector consists of a neural network $f_{ad} : X_i \rightarrow \{P_i^{(c)}\}^{c \in \mathcal{C}}$ mapping the input data $X_i$ to the probability of $X_i$ belonging to each category. 
$P_i^{(c)}$ is the probability of $X_i$ belonging to category $c$.
The category with the highest probability is chosen as the label for instance. 
Training the detector seeks to minimize the cross-entropy loss:
\begin{equation}
    L_{CE} = - \sum_{i=1}^{N} \sum_{c \in \mathcal{C}} 1\{\ell_i = c\} \log \left( P_i^{(c)} \right)
\end{equation}

Again, we develop an LSTM neural network-based attack detector (see in Table~\ref{table:lstmNNarchitectures} in the Appendix) given its suitability for time-series and for comparison between the supervised and unsupervised attack detection methods.

{To deploy the anomaly and attack detectors, the generation's governor intelligent electronic devices (IED) can be upgraded to collect measurements of the governor control signal and local frequency. The IED will store the last $n_{i-1}$ measurement samples, add the latest sample, and perform the neural network computations to classify the system data. The classification can be communicated to the grid operator SCADA system to alert on attacks, or actions can be programmed into the IED to autonomously mitigate detected attacks.}

    
    
        
        

\section{Results} \label{sec:results}

\begin{figure}
    \centering
    \includegraphics[width=\columnwidth]{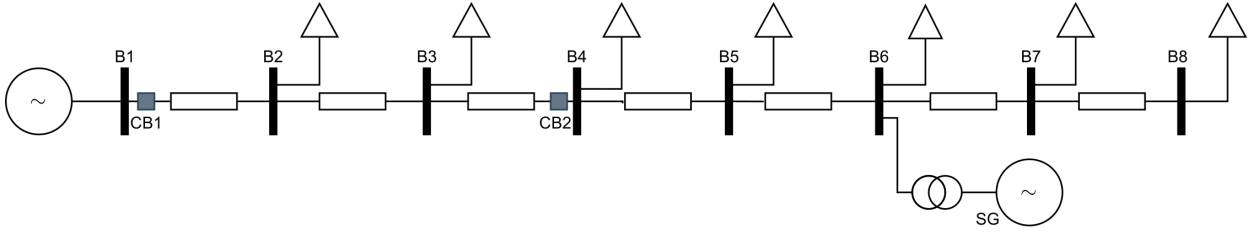}
    \caption{Microgrid testbed. SG is the synchronous generator which is target for the attacks.}
    \label{fig:mgTestbed}
\end{figure}

This section demonstrates RL's use for FDI and load switching attacks synthesis, defense using the attack and anomaly detectors, benefits of the RL method, optimality evaluation of RL attacks, and limitations and challenges of the method. 
Our results were validated using three different microgrid models (MG1 to MG3), with the same network as depicted in Figure \ref{fig:mgTestbed}, but different { load frequency control} parameters, as listed in Table \ref{table:systemsParameters} in the Appendix.
MG1 is the base microgrid model and replicates a 2.5 MVA rated microgrid in rural Ontario, Canada.
Microgrids, with their low inertia, are vulnerable to cyberattacks, making them ideal for studying power system vulnerabilities. 
FDI attacks compromise the control of the synchronous generator (SG in Fig. \ref{fig:mgTestbed}) and the load switching attacks compromise the switching of one or more of the microgrid's loads.
RL training employs a simplified LFC model of the microgrid, and is verified using time-domain simulations on a detailed model built using MATLAB Simscape.
MATLAB and Simulink were employed to generate the results.

\subsection{False data injection}

\begin{figure}
    \centering
    \includegraphics[width=0.5\columnwidth,trim={0 0 0 0},clip]{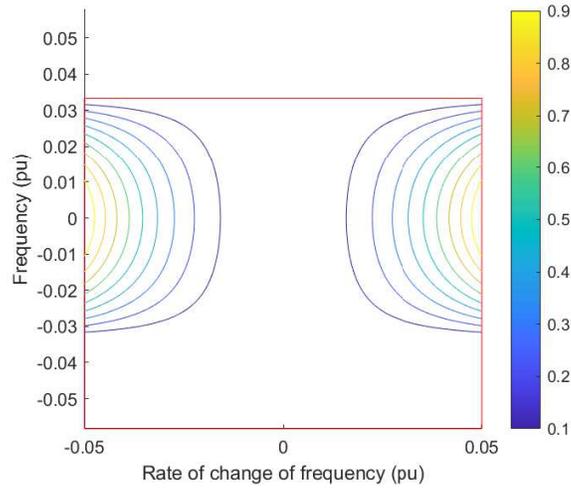}
    \caption{False data injection attacks reward function. The agent is rewarded for increasing the rate of change of frequency while maintaining small frequency deviation.}
    \label{fig:reward_fdi}
\end{figure}

\begin{figure}
    \centering
    \includegraphics[width=0.6\columnwidth,trim={0 0 0 0},clip]{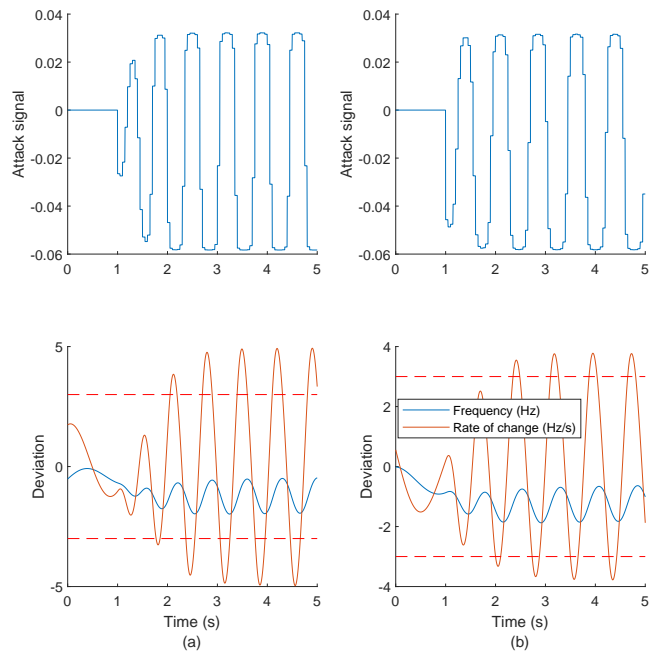}
    \caption{Detailed model simulations: Demonstrating the ability of the RL agent to attack different systems. Each column corresponds to a different system -- with eigenmodes located at (left: MG2) 4 and (right: MG3) 3.4 rad/s. The RL agent is able to adapt its attack.}
    \label{fig:ddpgAttacks}
\end{figure}

Recall that we employ a DDPG RL agent to inject false data into any of the inputs $\{p_1, p_2, p_3\}$ of a single-area system while observing the frequency and its rate of change.
The following results are specific to corrupting the frequency measurement to the control center ($p_1$); but are generally applicable to the other inputs.
Following experimentation, we list the following decisions we made in the design of the RL agent:
\begin{itemize}
    \item We bound the RL action space, representing the injected frequency bias, to $[-0.1, 0.1]$ pu. The large action space makes it easier and faster for the RL agent to learn successful attack strategies and generate a larger variety of attacks. 
    Smaller bounds lead to longer convergence times during training. 
    After training, the RL action space can be scaled down to smaller, more practical bounds to destabilize vulnerable power systems. 
    For example, the RL agent's actions in the detailed model time-domain simulations (Fig. \ref{fig:ddpgAttacks}) are restricted to the range $[-3.5, 2]/60$ pu to match the frequency range ($56.5$ to $62$ Hz) in which the generator regulates the system frequency (refer to Table \ref{table:relaysettings}). 
    \item The large action space allows the agent to quickly discover simple bias attack to trigger UF or OF protection. 
    We employ reward function \eqref{eq:rewardFunction1} to encourage the agent to discover more complex attacks. The reward function is illustrated in Fig. \ref{fig:reward_fdi}. 
    The safety set $\mathcal{S}$, which the agent attempts to force the system to exit, is inside the area depicted by the red square.
    Function \eqref{eq:rewardFunction1} rewards the agent for increasing the rate of change of frequency towards and beyond the ROCOF relay setting while maintaining the frequency deviation small. 
    Additionally, the agent attains a high reward of $+20$ when the ROCOF relay trips and a high penalty of $-20$ when the either of the UF or OF relays trips.
    Without the penalization, the agent continues to prefer simple actions that trigger UF or OF protection. 
    Our experiments show that generally following the above guidelines facilitates RL training. 
    \item We limit each episode to 15 seconds to encourage the agent to destabilize the system quickly, and end the episode when the agent succeeds in triggering protection.
\end{itemize}

\begin{align} \label{eq:rewardFunction1}
    R_i &= \left( \frac{\hat{\dot{\omega}}}{0.05} \right)^2 \cdot \max \left( 0, 1 - \left( \frac{\hat{\omega}}{0.0\dot{3}} \right) ^2 \right) \\
    &+ 20 \{\hat{\dot{\omega}} \notin \mathcal{S} \} - 20 \{\hat{\omega} \notin \mathcal{S} \} \nonumber
\end{align}

The agent generates an oscillatory frequency bias to excite the mechanical eigenmode of the microgrid, leading to generation tripping in vulnerable microgrids. Fig. \ref{fig:ddpgAttacks} shows (detailed-model) time-domain simulations of two vulnerable microgrid testbeds, each with a different eigenmode frequency. The top plot in each column depicts the injected attack signal, with the frequency and rate of change of frequency deviation plot below it. The horizontal red dashed lines indicate the ROCOF protection relay bounds, which trigger the corresponding relay function and cause generation tripping when exceeded. For demonstration purposes, the relay triggering is suppressed to continue to demonstrate the RL agent's attack strategy. The agent successfully triggers the rate of change of frequency relay function in the systems.

Fig. \ref{fig:ddpgAttacks} demonstrates RL agents' adaptability. Being data-driven, they can easily adjust their attacks to destabilize different systems. The agent utilizes easily available system frequency measurements to tailor the frequency of its injected attack signal to the mechanical eigenmode of the system.

\subsection{Load switching}

\begin{figure}
    \centering
    \includegraphics[width=0.5\columnwidth,trim={0 0 0 0},clip]{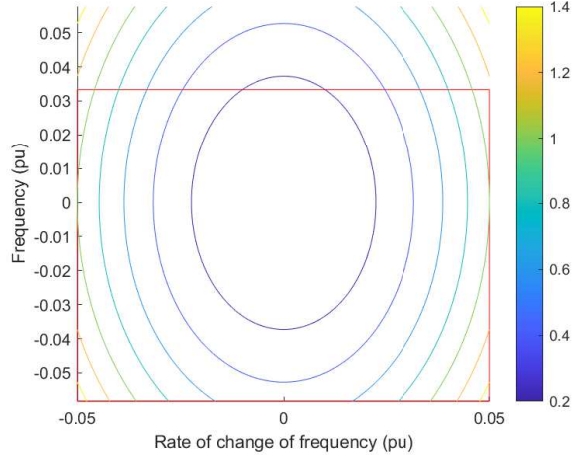}
    \caption{Load switching attack reward function. The agent is rewarded for increasing the frequency deviation and its rate of change towards the protection relay settings in red.}
    \label{fig:reward_loadswitch}
\end{figure}

\begin{figure}
    \centering
    \includegraphics[width=0.6\columnwidth,trim={0 0 0 0},clip]{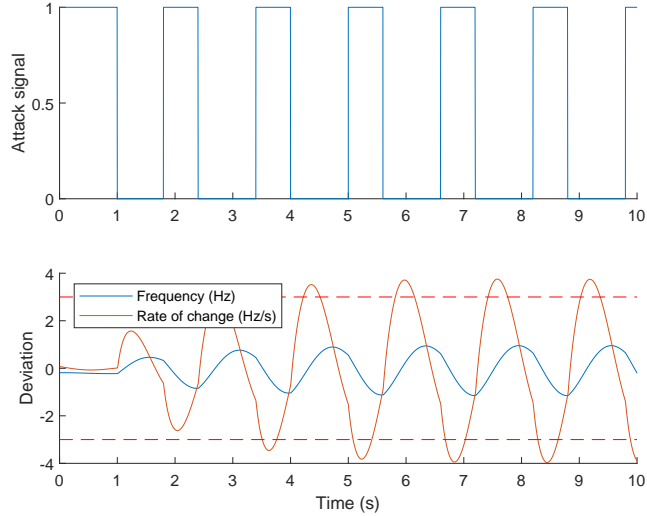}
    \caption{Detailed model simulation: Load switching attack against MG1. The load is switched on and off when the attack signal is 1 and 0, respectively. Generation is tripped when the rate of change of frequency (red in bottom plot) exceeds the ROCOF's relay settings (dashed).}
    \label{fig:loadswitch1}
\end{figure}

The RL agent learns to execute load switching attacks by manipulating the system load through $p_4$, while monitoring the frequency and its rate of change.
We use reward function \eqref{eq:rewardFunction2} to incentivize the agent to increase the frequency or rate of change of frequency deviations, with high rewards of $+20$ earned when any of the UF, OF, or ROCOF relays trip. 
The reward function is illustrated in Fig. \ref{fig:reward_loadswitch}.
The change in the reward function (from \eqref{eq:rewardFunction1}) is attributed to the difficulty of tripping UF/OF protection with switching attacks.

\begin{equation} \label{eq:rewardFunction2}
    R_i = \left( \frac{\hat{\dot{\omega}}}{0.05} \right)^2 + \left( \frac{\hat{\omega}}{0.08\dot{3}} \right)^2 + 20 \{(\hat{\omega}, \hat{\dot{\omega}}) \notin \mathcal{S} \}
\end{equation}

Fig. \ref{fig:loadswitch1} shows (detailed-model) time-domain simulations of a fixed load switching attack, wherein 464 kW (0.18 pu) of MG1's load is switched on (1) and off (0), thereby exciting the mechanical eigenmode and eventually tripping the generation.
We present examples of aggregate load switching attacks later in Fig. \ref{fig:duringtraining2}.

\subsection{Supervised attack detection}

\begin{figure}
    \centering
    \includegraphics[width=0.8\columnwidth,trim={0 0 0 0},clip]{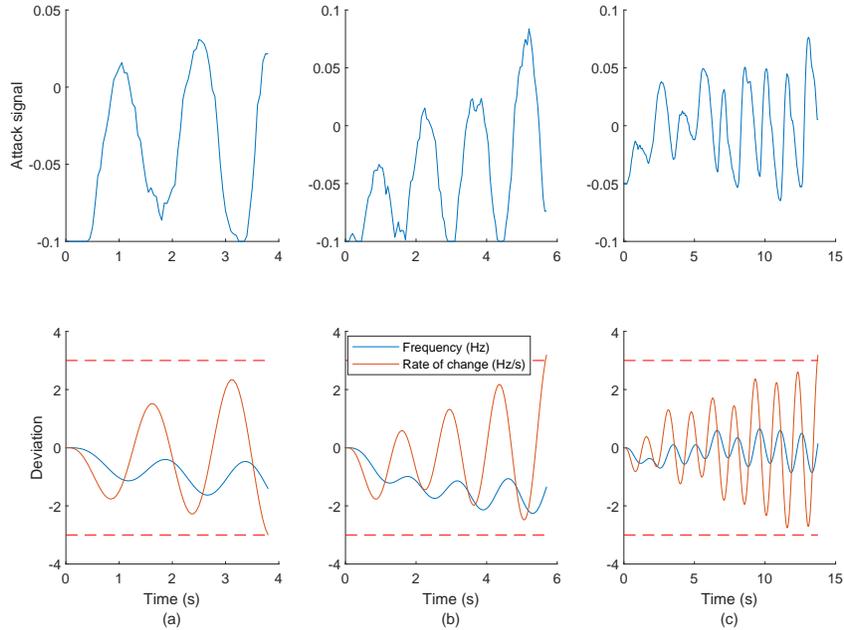}
    \caption{Different FDI attacks against MG2. The attack signal in the top figure is the per-unit injected frequency bias.}
    \label{fig:duringtraining1}
\end{figure}

\begin{figure}
    \centering
    \includegraphics[width=0.8\columnwidth,trim={0 0 0 0},clip]{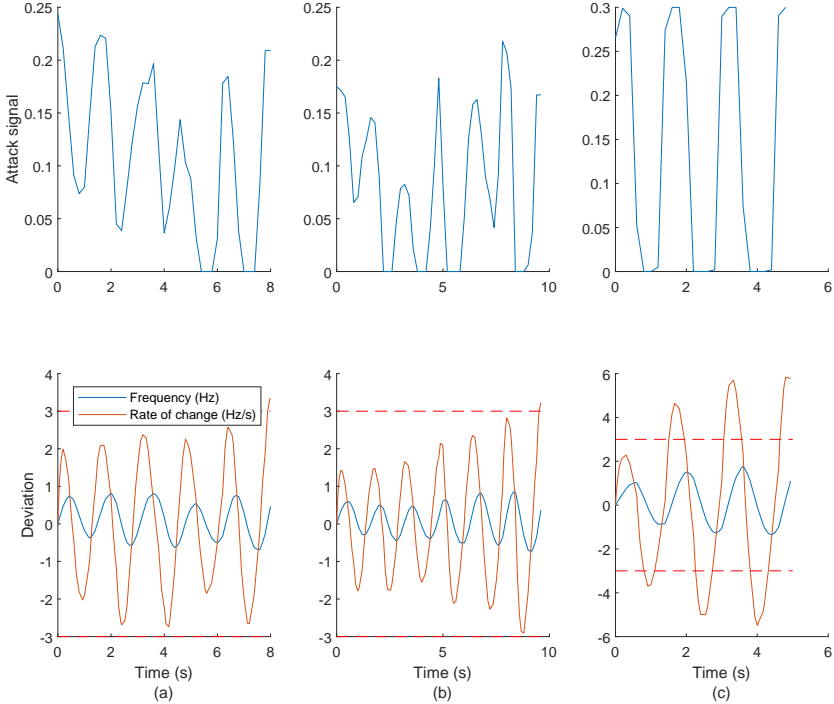}
    \caption{Different aggregate load switching attacks against MG1. The compromised load capacity load is 0.3 pu. The attack signal in the top figure is the amount of load that is switched off.}
    \label{fig:duringtraining2}
\end{figure}

In this section, the RL agent is trained to generate a large attack dataset to train the supervised-learning attack detector. To generate the data, the agent is trained with more emphasis on increasing exploration and delaying learning convergence. Figs. \ref{fig:duringtraining1} and \ref{fig:duringtraining2} showcase a variety of frequency measurement corruption and load switching attacks, respectively, collected during RL training that destabilized the LFC model. Although these attacks are generally not optimal (e.g., in terms of time-to-failure), they are still able to trigger protection relays and therefore warrant attention.

\begin{figure}
    \centering
    \vspace{2mm}\includegraphics[width=0.6\columnwidth]{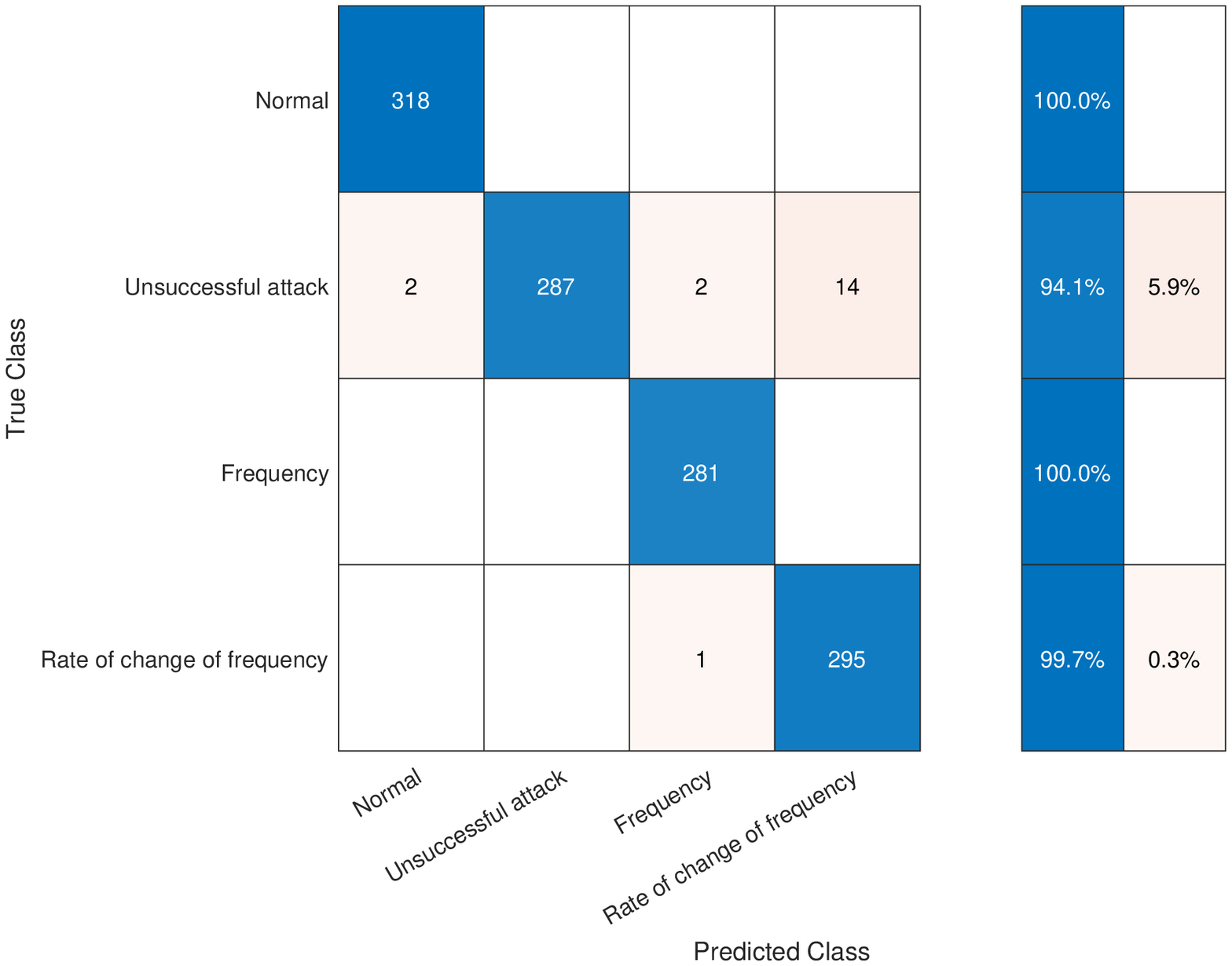}
    \caption{Supervised attack detector confusion matrix. The attack detector has a classification accuracy of 98.4\%}
    \label{fig:confMatrix}
\end{figure}

We gather 4000 records in the dataset, equally distributed among the 4 categories outlined in Section \ref{subsec:supervisedDetector}, and allocate 15\% and 30\% of the data for validation and testing, respectively. The detector's performance on the test data is presented in the confusion matrix in Fig. \ref{fig:confMatrix}, with an accuracy of 98.4\%.

\subsection{Unsupervised anomaly detection}

\begin{figure}
    \centering
    \includegraphics[width=0.6\columnwidth,trim={0 0.4cm 0 0},clip]{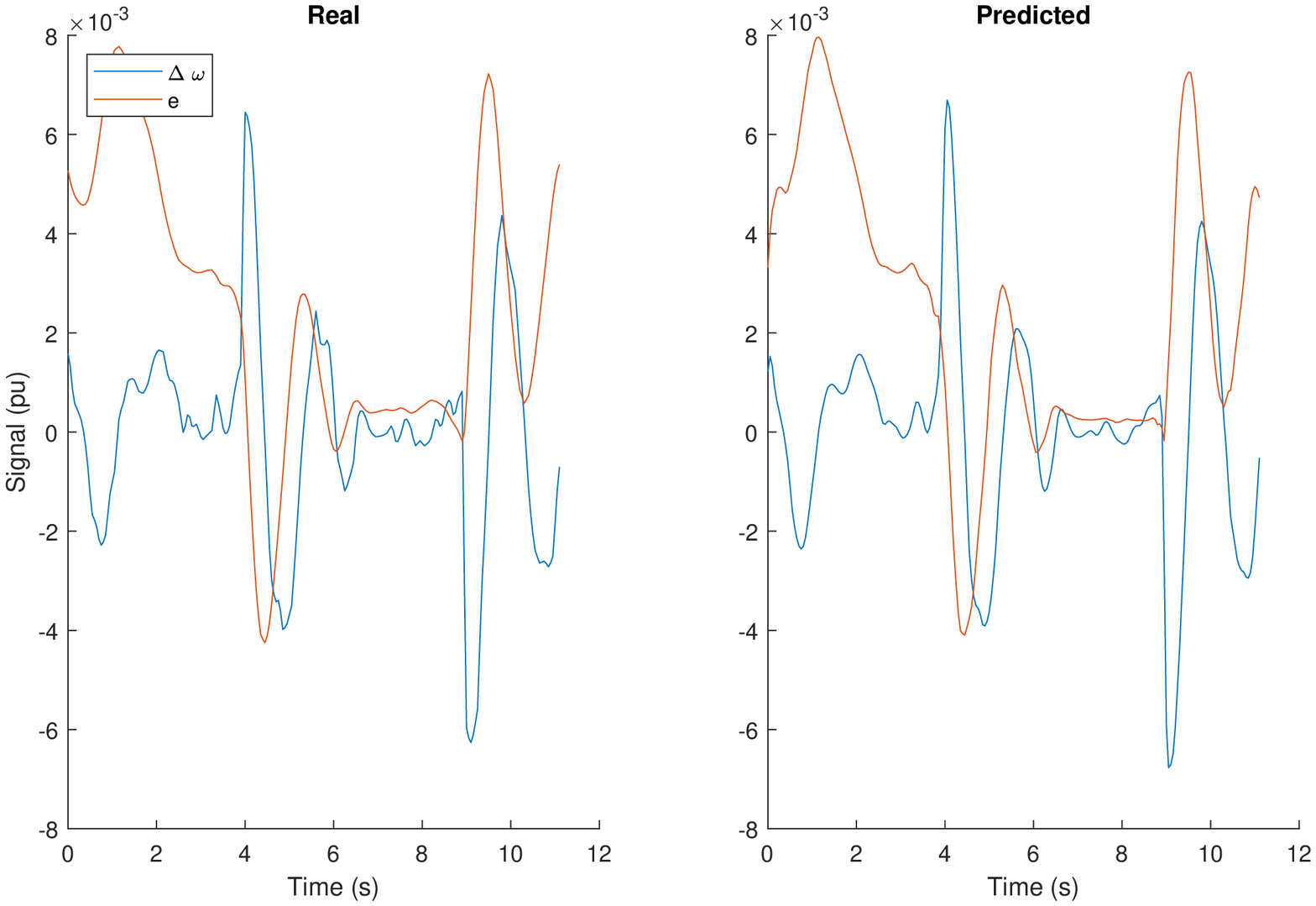}
    \caption{Autoencoder data reconstruction on normal operation data. Observe the high reconstruction similarity when the input data (left) is similar to the training data of the autoencoder.}
    \label{fig:autoencoderPrediction}
\end{figure}

\begin{figure}
    \centering
    \includegraphics[width=0.6\columnwidth]{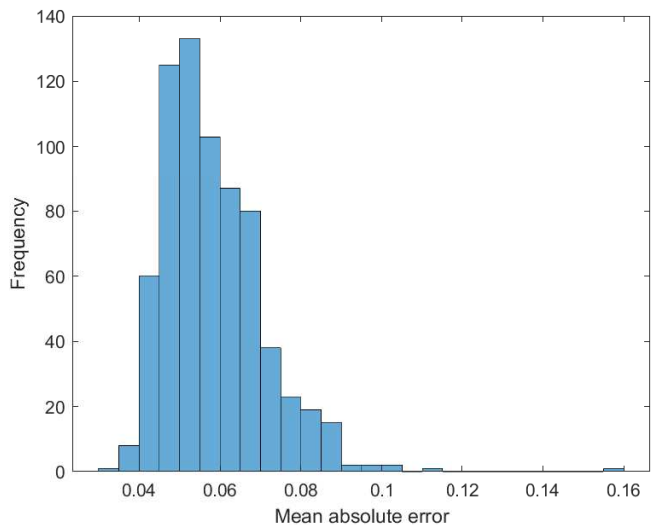}
    \caption{Histogram of the autoencoder's validation data mean absolute error. Higher errors during testing indicate anomalous data.}
    \label{fig:autoencoderMAE}
\end{figure}

Fig. \ref{fig:autoencoderPrediction} demonstrates the reconstruction of the trained autoencoder. 
The left plot shows a portion of control signal and frequency measurement collected during normal operation. 
The right plot shows the autoencoder's reconstruction of the data. 
By learning the patterns of data during normal operation, the autoencoder is able to reconstruct the data with low error.
Fig. \ref{fig:autoencoderMAE} is a frequency histogram of the autoencoder's reconstruction mean absolute error on the validation data. For comparison purposes, the {test and validation sets are} the same as that used to compute the supervised detectors' accuracy.
Based on the histogram, we select an error value of $0.2$ to classify between normal and anomalous behavior, which yields $100\%$ accuracy classifying between normal and attack behavior. 

When comparing the unsupervised anomaly detector's accuracy to that of the supervised attack detector in classifying normal and anomalous (comprising successful and unsuccessful attacks) operation, the detectors are comparable --  at 100\% and 99.8\%, respectively.
If, however, we compare their accuracy in classifying behavior preceding relay triggering (comprising successful attacks) and behavior that is not (comprising normal operation and unsuccessful attacks), the anomaly detector's accuracy is 74.5\% compared to 98.6\%.
{This echoes a major drawback of unsupervised methods, which is their high false alarm rate for safe anomalous events that are difficult to exhaustively include in their training data, even when these events do not have any impact on the system.}
We discuss consequent implication and opportunities further in the discussion.

\subsection{Discussion} \label{subsec:discussion}

\subsubsection{{RL suitability}}

{
We demonstrated that RL can be used to compromise LFC. 
Offensively, an attacker can execute the RL actions in a man-in-the-middle attack or package the RL as a malicious software. 
The RL agent offers a simple, fast, flexible, and adaptive approach to cyber offense, modifying its actions to attack different systems without the need for prior reconnaissance to collect system information or exact models of the targeted systems. 
This alerts to the urgency of using RL defensively, as suggested by the paper, to proactively identity and collect attack strategies before system vulnerabilities are exploited.
}

\subsubsection{Attack model assumptions}

The attacker's knowledge, resources and limitations are specified as part of developing the RL problem. 
For example, in the presented attacks, we specify that the attacker has no prior knowledge, their disclosure resources enable them to observe the grid frequency, and their disruption resources enables them to compromise specific communication channels (outlined in Section \ref{sec:threatModel}). 
Their limitations are formulated in the constraints on the action space.
If the learning converges, the specified attack and threat models are sufficient to execute the attack.

This can enable verification of assumptions on the attack/threat model.
For example, while presenting switching attacks, the authors in~\cite{hammad2017class} assume that the attacker needs prior knowledge of the inter-area eigenmodes to execute the attack.
In a preceding reconnaissance attack, the authors suggest that the attacker can impose faults at the compromised load (e.g., by employing a built-in de-energization circuit breaker). The attacker records transient line measurements following the fault. Next, the attacker performs spectral independent component analysis (ICA) analysis on the recorded data to reveal the inter-area modes.
Evident from results in Fig. \ref{fig:ddpgAttacks} and \ref{fig:loadswitch1}, we show that a pre-trained RL agent does not necessarily need this knowledge; but only needs to observe the frequency during attack execution -- alarming us that the attack can be executed with relative ease.

\subsubsection{System vulnerability testing}

We demonstrated that our method can be used to synthesize multiple attacks against a system during the RL training process.
Practically, this yields a system vulnerability testing application for our research.
For example, the RL training can be performed on a offline system model. 
Simulations can reveal vulnerabilities that need to be patched before deployment.
Additionally, preceding every system change or upgrade, RL training can reveal vulnerabilities before cyberattacker capitalize on them. 

Our method can also be used to validate defense strategies. 
After a vulnerability is identified in training, defense methods including upgrading control algorithms (physical), upgrading code security (computational), or adding channel redundancy (communication-based) can be designed and incorporated into the offline system model. 
If the defense method passes the  previously successful logged attacks and further RL training without any system failures, then the defense can be deployed to enhance system security.

We also demonstrated that a single RL agent can execute successful attacks against different systems. 
This provides an opportunity to collect an `arsenal' of RL agents and provide them for system owner-operators to automatically test vulnerabilities. 
After modelling their system, the owner-operator can retrieve the RL agents from repositories and have each RL agent check for a specific system vulnerability. 
In this way, our method can transform cyber-physical vulnerability scanning to an approach that is similar to current cyber malware scanning.

\subsubsection{Attack optimality}

\begin{figure}
    \centering
    \includegraphics[width=0.5\columnwidth]{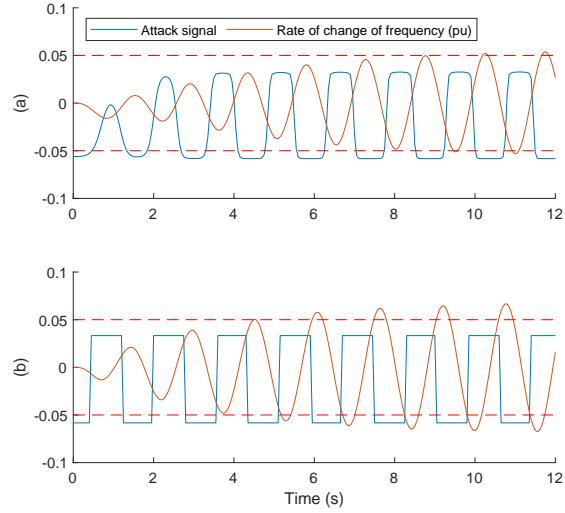}
    \caption{Comparing the RL convergence to the optimal attack. (a) shows the optimal attack with attack frequency matching the oscillatory eigenmode frequency. Note that this requires prior system knowledge. (b) is the RL-generated attack.}
    \label{fig:optimizationComparison}
\end{figure}

If we quantify attack optimality by (shortest) time-to-failure or (maximum) deviation of frequency (or rate of change of frequency), then the RL method only provides sub-optimal attack policy.
The top plot of Fig. \ref{fig:optimizationComparison} shows an optimal FDI attack with a frequency matching the oscillatory eigenmode of the system. The bottom plot shows the RL generated attack signal following learning convergence. 
Despite being very similar, the optimal attack still produces a larger rate of change of frequency deviation and triggers protection faster. Both attacks still successfully trigger protection.

We provide the following explanation to why the RL method might not yield an optimal policy.
Once the RL agent successfully triggers a protection relay function, the episode stops. 
The largest reward is achieved by triggering the rate of change of frequency function.
Policy improvement occurs over episodes as the agent seeks the small rewards gained by inducing higher frequency and rate of change frequency deviation (refer to (\ref{eq:rewardFunction1})).
Hence, it is likely that the training will stop at a sub-optimal stage. 
{More} episodes will be needed to reach optimality.

\subsubsection{Integrated attack detection}

\begin{figure}
    \centering
    \includegraphics[width=0.8\columnwidth]{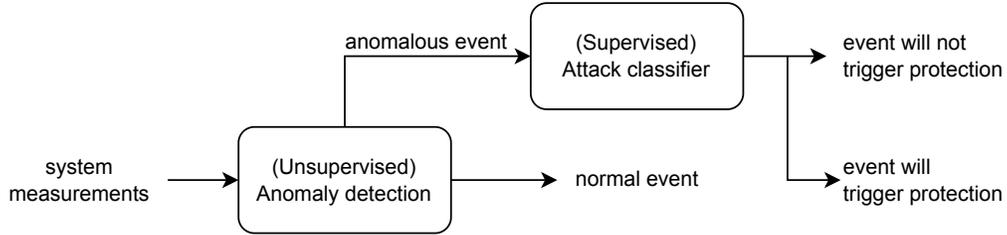}
    \caption{Integrated defense block diagram. The supervised attack detector improves defense by avoiding unnecessary tripping.}
    \label{fig:integratedDef}
\end{figure}

False triggers due to rare normal events is a significant shortcoming of unsupervised anomaly detection methods. 
In our study, the unsuccessful attacks category represents events that are not followed by any loss of generation due to false relay triggering. 
The supervised attack detector successfully classified 94\% of these instances.
The anomaly detector's training, however, does not enable it to recognize these events.
{Consequently, the supervised attack detector is better equipped to alert only to malicious cyberattacks, while suppressing unnecessary responses to safe anomalous events.}
In practice, an integrated approach can be used to detect attacks -- capitalizing on the anomaly detector's strength in detecting normal behavior, and the attack detector's strength in detecting attacks that trigger protection. 
Fig. \ref{fig:integratedDef} demonstrates the integrated approach. 
Events that are classified as normal by the anomaly detector are accepted, while those that are not are evaluated by the attack detector to determine if they will result in relay triggering and require corrective action.
The attack detection accuracy of this approach is 98.6\%.

\subsubsection{Limitations and challenges}

RL training is time-consuming and computational intensive, which only increases for more complex environments.
In this paper, we used linearized models to speed training. 
Note that the use of linearized models is widely acceptable given that grid cyber-physical security studies utilize simplified models for ease of mathematical formulation. The RL agents still successfully destabilized the detailed microgrid models.

On another note: as training progresses and learning converges, the agent is taking more exploitative actions. The variation between the generated attacks becomes smaller as training progresses. 
The distinct attacks that are generated early in training when the agent is explorative provide important insight and data samples for the attack detectors training. 
As such, it is beneficial to prolong agent exploration, which unfortunately increases time to learning convergence and the number of episodes.

The RL training does not produce a comprehensive dataset of all attacks that cause system failure. 
However, with the dataset generated during RL training, security engineers can develop valuable defense strategies.

\section{Conclusion} \label{sec:conclusion}

Proactively identifying grid vulnerabilities and attack strategies is critical to anticipate attacks and patch grid weaknesses before they are exploited.
We develop deep (DDPG) RL agents to execute false data injection and load switching attacks against LFC. 
The RL-generated attacks directly induce protection relay tripping and generation loss, which can subsequently lead to grid instability and blackout.
The process of training the RL agent provides valuable insight into attacker resources and strategies, including specifying attack and threat models and generating attack datasets. 
The attack datasets can be used defensively to inform, evaluate and develop defense strategies.
We develop an LSTM-based supervised-learning model to classify and detect attacks and compare it with state-of-the-art autoencoder-based anomaly detection.
The supervised attack detector achieves comparable accuracy (99.8\%) when classifying normal and anomalous operation.
While anomaly detection is not equipped to identify anomalous events that do not induce relay triggering (such as normal rare events), the supervised attack detector classifies these events with high accuracy (98.6\%).
We propose an integrated attack detector that capitalizes on the strengths of anomaly detection and supervised attack detection to improve attack detection accuracy while reducing false detection.

\appendix

The state matrices in (\ref{eq:stateSpace}) are as follows:
{\footnotesize
\begin{align} 
    \bm{\Dot{x}} =& 
    \begin{bmatrix} 0 & 0 & 0 & -(k B) & 0 & 0\\
    1/\tau_G & -1/\tau_G & 0 & -d/ (\tau_G) & 0 & 0\\
    0 & 1/\tau_T & -1/\tau_T & 0 & 0 & 0\\
    0 & 0 & 1/M & -D/M & 0 & 0\\
    0 & 0 & 0 & 1/\tau_\omega & -1/\tau_\omega & 0\\
    0 & 0 & 1/(M \tau_\nu) & -D/(M\tau_\nu) & 0 & -1/\tau_\nu
    \end{bmatrix} \bm{x} \nonumber\\
    +& \begin{bmatrix} 0 & 0\\
    0 & -k\\
    0 & 0\\
    -1/M & 0\\
    0 & 0\\
    -1/(M \tau_\nu) & 0
    \end{bmatrix} \bm{u} + 
    \begin{bmatrix} -(k B) & k & -k & 0\\
    0 & 0 & 0 & 0\\
    0 & 0 & 0 & 0\\
    0 & 0 & 0 & -1/M\\
    0 & 0 & 0 & 0\\
    0 & 0 & 0 & -1/(M \tau_\nu)
    \end{bmatrix} \bm{p}
\end{align}
}

\begin{table}[H]
\scriptsize
\centering
\caption{System Data}
\begin{tabular}[t]{lcccc}
\toprule
\textbf{Parameter} & \textbf{Symbol} & \textbf{MG1} & \textbf{MG2} & \textbf{MG3}\\
\midrule
AGC gain & $k$ & 3 & 10 & 12 \\
Droop gain & $d$ & 40 & 50 & 50 \\
Governor time-constant & $\tau_G$ & 0.08 & 0.08 & 0.1 \\
Turbine time-constant & $\tau_T$ & 0.45 & 0.45 & 0.45 \\
Generator inertia & $M$ & 6 & 6 & 8 \\
Damping constant & $D$ & 0.03 & 0.03 & 0.03 \\
\bottomrule
\label{table:systemsParameters}
\end{tabular}

\begin{tabular}{lc}
     Frequency sensors time-constants & $\tau_\omega = \tau_\nu = 0.1$\\
     Control center frequency measurement gain & $B = 1$ 
\end{tabular}

\end{table}

\begin{table}[H]
\scriptsize
\centering
\caption{DDPG neural network architectures}
\begin{tabular}[t]{lclc} 
\toprule
\textbf{Actor network}\\
\midrule
\textbf{Layer} & \textbf{\# of units} & \textbf{Hyperparameters}\\
\midrule
Input & 2 $(\Delta \hat{\omega}, \hat{\dot{\omega}})$ & $M = 128$\\
Normalization & 2 & $\alpha_\theta = 10^{-4}, 
 \alpha_\phi = 10^{-3}$\\
Fully-connected & 100 & $\gamma = 0.99$\\
ReLU &  & $\tau = 10^{-3}$\\
Fully-connected & 50 & $N \sim \mathcal{N}(0, 0.3)$\\
ReLU\\
Tanh (or Sigmoid)\\
Scaling & 1\\
Output & 1 ($A$)\\
\midrule
\textbf{Critic network}\\
\midrule
\textbf{Layer} & \textbf{\# of units} & \textbf{Layer} & \textbf{\# of units}\\
\midrule
Input & 2 $(\Delta \hat{\omega}, \hat{\dot{\omega}})$ & Input & 1 ($A$)\\
Normalization & 2 & Normalization & 1\\
Fully-connected & 100 & Fully-connected & 50\\
ReLU\\
Fully-connected & 50\\
Addition & 50 & $\swarrow$\\
ReLU\\
Fully-connected & 1\\
Output & 1 $Q(\Delta \hat{\omega}, \hat{\dot{\omega}}, A)$\\
\bottomrule
\label{table:ddpgNNarchitectures}
\end{tabular}
\end{table}

\begin{table}[H]
\scriptsize
\centering
\caption{Detectors neural network architecture}
\begin{tabular}[t]{lc|lc} 
\toprule
\textbf{Supervised} & & \textbf{Unsupervised}\\
\midrule
\textbf{Layer} & \textbf{\# of units} & \textbf{Layer} & \textbf{\# of units}\\
\midrule
Sequence input & ($\Delta e, \Delta \hat{\omega}$) & Sequence input & ($\Delta e, \Delta \hat{\omega}$)\\
LSTM & 75 & BiLSTM (w/ normalization) & 36\\
Dropout ($10\%$) & & ReLU\\
LSTM & 50 & BiLSTM (w/ normalization) & 8\\
Dropout ($20\%$) & &  ReLU\\
LSTM & 35 & BiLSTM (w/ normalization) & 36\\
Dropout ($10\%$) & ReLU & Sequence output\\
Fully-connected & 4\\
Softmax & 4 ($c \in \mathcal{C}$)\\
\bottomrule
\label{table:lstmNNarchitectures}
\end{tabular}
\end{table}

\bibliography{references}

\end{document}